# Semantic Validity vs. Graphical Arrangement in Visualization Grammars

Cynthia A. Huang [1]
LMU Munich, Munich Center for Machine Learning (MCML)

Matthew Kay [2]
Northwestern University

## Abstract

Systems for modular, declarative and compositional specification for data visualizations, otherwise known as visualization grammars, often claim Wilkinson's Grammar of Graphics as motivation, inspiration or foundation. However, despite the shared intellectual lineage, visualization grammars can vary greatly in design scope and focus. In particular, systems that strongly adhere to domain-specific notions of correctness must trade-off expressivity for validity. This trade-off between semantics and graphical flexibility is currently under-articulated, leading to difficulty integrating various GoG-inspired visualization contributions, especially between InfoVis and Statistical Graphics. We propose three new types of visualization contributions to support such integration: *Semantic Validity Theories*, *Semantic Visualization Systems* and *Graphical Arrangement Systems*. The first two prioritize the adherence of data visualizations to correctness conditions, such as statistical or mathematical transformations and data semantics, whilst the third largely leave validity concerns to the user. We illustrate these new contribution types using examples from two recent grammar of graphics inspired papers, and show how they supported more nuanced notions of expressivity and usability in evaluation and design of visualization systems.

## 1 Introduction

Two recent papers in premier venues in InfoVis and Statistical Graphics both study the same chart type, cite Wilkinson's grammar of graphics [24], and invoke the same design goal of *expressivity* [2, 16]. Both use stacked bar charts as a motivating problem and running example. Both papers identify and address limitations of current grammars. Both seek to ensure adherence to Gestalt principles in resultant data visualizations. However, one proves that stacking group means is statistically invalid, while the other provides an operator that makes stacking group means trivially easy – suggesting conflicting or at least incommensurate contributions. Despite overlapping motivations, and shared intellectual lineage, we argue these two contributions cannot be, and *should not* be directly compared.

Pollock and Satyanarayan [16] propose a "grammar of more graphics". The authors claim that GoFish overcomes consistency and composability limitations in existing GoG implementations. They achieve this by operationalizing Gestalt grouping principles as composable graphical operators to achieve parsimonious transformation between specifications of these seemingly unrelated graphical arrangements. As an example, they recast stacked and dodged bar charts, along with ribbon charts and polar ribbon charts as variations of the simple bar chart to show how their grammar overcomes the "sharp *expressivity* cliffs" of existing grammars.

Bartonicek et al. [2] examine the widespread usage of stacked bar charts over dodged bar charts in interactive graphics, and offer an explanation for this preference based in Gestalt principles. They then proceed to argue that only statistics with monoidal structure (e.g. counts, sums, products) should be displayed as stacked bars with linked highlighting of single group subsets (parts of whole), based on the algebraic properties of summary functions. The authors offer this result, and a similar one for multi-group selection, as a preliminary contribution towards complete algebraic foundations for 'coherent and well-behaved' interactive graphics. They invoke a more constrained notion of *expressivity* that is conditional on statistical and mathematical validity. In fact, Bartonicek et al. [2] explicitly caution against treating stacking as a purely graphical operation – a view that appears to directly challenge the validity and usefulness of the stacking operator in GoFish.

We argue the conflict reflects previously unarticulated distinctions in design goals and priorities between different grammar-based visualization systems. To articulate these distinctions, we propose the concepts of *semantic fidelity and validity*, *graphically tractable data*, and *semantic and graphical expressivity* and characterize three new archetypal visualization contribution types, *Semantic Validity Theories*, *Semantic Visualization Systems* and *Graphical Arrangement Systems*. We then demonstrate the utility of this new vocabulary supporting the description, evaluation and integration of parallel and complementary GoG-inspired research across visualisation research fields.

## 2 Motivating contributions

To motivate our proposed vocabulary, we describe the two bar-chart related contributions in detail, and surface the latent hazard that emerges when they are brought into contact.

Bartonicek et al. [2] establish a perceptual argument for semantic fidelity in interactive bar charts. They propose that stacking is more appropriate than dodging for displaying linked selection, because the four Gestalt principles of proximity, similarity, closure, and common region ensure that a stacked bar presents as a unified visual entity whose outline remains stable across all interaction states. They further argue that validity of static graphics does not necessarily imply that interactive sequences of those graphics (e.g. highlighted bar charts) will also be semantically valid. Then, via algebraic analysis, they derive formal admissibility conditions: statistics must be monoidal (associative with a neutral element) for valid single-group linked selection, and must additionally have an inverse — forming a group — for valid multi-group linked selection. Although Bartonicek also implements a prototype system [1] for creating valid interactive bar charts, their primary contribution is the validity theory itself.

GoFish [16] separates spatial arrangement from mark definition for the goal of greater expressivity. They begin by observing that GoG systems bundle shape and layout into composite marks, often producing one custom mark type per arrangement. By contrast, GoFish reduces marks to primitive shapes and formalizes arrangement as three composable Gestalt operators: Stack (uniform spacing), Connect (element connectedness), and Enclose (common region). These operators can be nested arbitrarily and can share children across layouts, allowing stacked and grouped bar charts, waffle, ribbon, and polar ribbon charts to all be expressed as variations of a basic bar chart specification rather than as bespoke composite marks. GoFish handles the resulting computational challenges including inferring how shapes should behave under coordinate transforms, and resolving size scales across unbounded nesting depth declaratively, without exposing this machinery to the user. GoFish's specific advance is thus architectural: it extends the graphical arrangement design space by making recursive spatial composition tractable.

In 'untangling GoG marks from operators and coords', Pollock and Satyanarayan [16] decouple the graphical arrangement task from underlying statistical semantics. The current implementation of GoFish is restricted to static graphics, so the consequences of this decoupling are limited to possible silent errors in static graphics, but carry no immediate interactive consequences. However, the authors identify interactivity as a future direction. Unfortunately, if linked selection was added without integrating validity considerations, the grammar would silently permit combinations, such as stacking of group means

[1] cynthia.huang\@lmu.de
[2] matthew.kay\@u.northwestern.edu

(via `StackX()`), that Bartonicek et al. [2] prove are mathematically meaningless, and statistically invalid. Although, it is not clear that visualization systems should always strictly prevent the creation of 'mathematically invalid' interactive graphics, at minimum, such risks warrant careful consideration. The conceptual vocabulary proposed in this paper helps us better characterize, and navigate such nuances. Before introducing that vocabulary, we situate our work within existing formalizations of validity in visualization.

## 3 Background

### 3.1 Validity in Visualization

There is extensive prior work formalizing notions of validity, and documenting constraints arising from these notions. Kindlmann and Scheidegger's [8] Algebraic Visualization Design establishes that mathematical structures in data should correspond to mathematical structures in perception; McNutt et al. [12] identify visualization mirages — silent, significant errors arising from broken connections between data and visual form. Similar notions have been used to define (for example) *correctness* for probabilistic visualizations [7, 18, 23], structure-preserving scale transformations [10], and distance-preserving visual embeddings [4, 5]. These contributions treat fidelity and faithfulness to data-driven or domain-specific semantics as formalized concerns to be considered alongside perceptual and graphic design principles. Prior commentary on these formalization and correctness concerns can also be found in surveys and comparisons of visualization systems [9, 13, 19]. Our work particularly complements McNutt's [13] survey and analysis of JSON-style Domain-Specific Languages (DSLs) for visualization specification.

### 3.2 Validity in Grammar of Graphics Extensions

When introducing the *Grammar of Graphics*, Wilkinson insists that it should not be seen as a new graphics scripting language for statistical graphics, but rather a formal theory of grammatical rules for constructing an organized set of points (the 'graph') mathematically, and then using aesthetic mapping to represent the graph as a graphic. Statistical validity is an irreducible component of Wilkinson's GoG, whereby data transformations must remain coupled to visual representations, to prevent silent errors arising from breaking "the connection between the variables (i.e. data) and the graphics that represent them" [24p. 106]. This guarantee is about meaningfulness, not quality: Wilkinson notes that the grammar can still produce hideous graphics and does not prevent their misuse [24]. Constraints that rule out graphics the community deems ineffective or misleading come from downstream implementations, not from the grammar itself.

GoG-inspired implementations vary greatly in their adherence to Wilkinson's statistical validity principles. Wickham offers an alternative layered parameterisation of Wilkinson's GoG, and largely adheres to notions of statistical validity [21, 22]. The statistics layer in *ggplot2* explicitly places statistical transformations as a core component of *statistical graphics*[3]. Paired geom–stat defaults (e.g., `geom_bar` paired with `stat_count`) reflect valid statistics-to-visualization relationships, but the *ggplot2* also allows users to experiment with alternative geometries or statistics to create novel statistical graphics. In contrast, systems such as *Vega-Lite* [20], and *Encodable* [25] treat statistical transformations as a pre-processing step. Wu et al. [26] propose and formalize validity principles for a different domain: database visualization. They use the term *faithfulness* to refer to the requirement that underlying database constraints are preserved and accurately reflected in the resulting visualization.

## 4 Validity vs. Arrangement

Contributions like Bartonicek et al. that echo Wilkinson's validity concerns, generally treat domain semantics and associated correctness conditions as a first-class design concern. Others like GoFish prioritize graphical expressivity, and maximizing the space of graphical arrangements a user can reach. In this section we provide vocabulary for describing this divergence.

The concepts of semantic fidelity, reduction and graphical arrangements allow us to characterize two distinct design and implementation orientations in visualization systems: *semantic visualization* and *graphical arrangement*. These distinctions allow us to better characterize, distinguish and integrate grammar-based visualization contributions.

### 4.1 Semantic Fidelity and Domain Semantics

Every data domain has mathematical and logical formalizations which define what its objects, operations, and relationships mean. These formal meaning structures, or **domain semantics**, are well-established for traditional domains such as space and time, and have long been incorporated into the design of computational tools — temporal and spatial database extensions, for instance, provide explicit support for temporally and spatially valid data operations. Extending this support into visualization systems requires defining valid graphical operations grounded in relevant domain semantics.

We use **semantic fidelity** to describe the degree to which a visualization accurately reflects the domain semantics of its underlying data. Fidelity is a graded property; a visualization may accurately represent some aspects of domain semantics but not others. For example, visualizations of time-series data may inaccurately represent the calendrical position of observations (e.g. months are ordered lexicographically), reflect partial ordinal structure of the underlying data (days within the month). Such losses in fidelity often begin with semantic reduction at the data level, whereby visually relevant semantics are lost – e.g. by encoding dates as numerics without calendrical information.

We further define **semantically valid visualizations** as visualizations that have sufficient fidelity for a particular usage context. **Semantic validity theories** are then formal frameworks that make domain semantics explicit and establish which graphical operations and arrangements are admissible. These validity frameworks also implicitly define semantically-constrained visualization design spaces. Establishing what operations and arrangements are admissible requires translation of mathematical and logical semantic structures into default encoding choices. Gestalt principles may also be incorporated to identify and compensate for interactions between the human perceptual system and the fidelity of underlying data semantics, as demonstrated in Bartonicek et al.'s recommendation of stacking over dodging for linked selection in bar charts (Section 2).

### 4.2 Producing semantically valid graphics

Semantic validity theories are valuable formalizations, but must be implemented to be useful in creating visualizations. Wilkinson's GoG is arguably *the* canonical semantic validity theory for statistical graphics, in that its structure ensures a graphic cannot be meaningless (though not that it is effective), and its implementation history illustrates this distinction. The syntax used by Wilkinson to specify grammatical visualizations was one of many possible interfaces or implementations. Wickham's layered grammar of graphics offered an alternative layered interface design and implementation of GoG's compositional object-orientated design [22]. Wickham's design has been widely adopted by and iterated on in the *ggplot2* R package and extension ecosystem. This sequence of developments illustrates how semantic validity theories are necessary pre-conditions for, but distinct contributions from graphics systems that can actually produce semantically valid graphics. Wu et al.'s [26] formalism and library for database visualization is a notable recent example of work that develops both semantic validity theory, and implements a visualization system. We refer to the latter as **semantic visualization systems**.

[3] We loosely use the term statistical graphics to refer to visualizations that explicitly adhere to statistical and mathematical notions of validity, and data visualizations or graphical arrangements for all other visualizations. The classification of statistical graphics usually cannot be made on the basis of the final visualization artefact alone, but is important for designing visualization systems that support workflows requiring statistical graphics (i.e. statistical analysis).

Semantic visualization systems must encompass the specification, construction, and rendering of visualizations which explicitly embed, validate, and preserve specific domain semantics required to accurately represent and interpret the underlying data being visualized. These requirements can be summarized in the following three core design principles:

1. **Semantic Validity**: ensure the semantic validity of user-specified graphics.
2. **Fluid Navigation**: minimize cognitive load and effort while iterating over valid graphics.
3. **Error Friction**: increase friction associated with producing graphics with invalid domain semantics.

### 4.3 Graphical Arrangements and Graphically Tractable Data

We contrast semantically-constrained design spaces and semantic validity theories with design spaces defined primarily by graphical arrangement. These design spaces reduce data, implicitly or explicitly, to representations directly manipulable by a graphical system, which we refer to as **graphically tractable data**. The degree to which graphically tractable data reduces or retains domain semantics depends on the graphical system. Furthermore, the semantic validity of graphics produced by arranging these semantically reduced data is not guaranteed, though nor is it precluded. The validity of graphics in unconstrained graphical arrangement space is orthogonal to the combinatorial extent of the design space, as the connection between the original domain-semantic object and the graphical form is not explicitly considered.

Although the underlying domain semantics are generally not considered when enumerating graphical arrangements, perceptual factors such as those described by Gestalt principles can serve as an alternative constraining lens over visualization design spaces. For example, perceptual grouping principles implicitly define ‘good’ or ‘effective’ graphical arrangements, a subset that forms a different type of constrained visualization design space. Pollock and Satyanarayan implement graphical operators and a spatial layout grammar over this perceptually constrained design space (Section 2).

### 4.4 Visualizing graphically tractable data

In contrast to semantic visualization systems, **graphical arrangement systems** are primarily concerned with computationally efficient rendering of a maximal range of distinct visual representations for graphically tractable data. Such systems generally do not validate or enforce the semantic validity of graphical arrangements, instead leaving that responsibility to the user or upstream data preparation tools. Consideration is however, often given to the usability of the system interface to navigate over graphical arrangements. This design orientation can be characterized by descriptive dimensions such as:

1. **Arrangement Coverage**: the range of distinct graphical arrangements expressible within the system
2. **Computational Efficiency**: time and resource required to render specified graphical arrangements
3. **Specification Viscosity**: the user effort required to reach graphical arrangement variants from an existing specification.

## 5 Diagnosing the divide

Table 1: Bartonicek et al. and GoFish compared as instances of different contributions. Both papers cite Wilkinson and use stacked bar charts as a running example, and their complementary relationship is clarified with our vocabulary.

| | Bartonicek et al. | GoFish |
|---|---|---|
| **Contribution** | Semantic Validity Theory | Graphical Arrangement System |
| **Primary goal** | Derive admissibility conditions | Maximise graphical expressivity |
| **Data** | Statistical semantics | Graphically tractable |
| **Semantics** | Monoidality and group conditions | Not enforced — decoupled from graphical arrangement |
| **Out of Scope** | Produce or render graphics | Determine whether a statistic is valid for a given arrangement |

We now revisit the contributions and latent hazard discussed in Section 2. Our proposed conceptual vocabulary helps clarify the relationship between the two bar-chart related contributions of Bartonicek et al. and GoFish. As summarized in Table 1, the former offers a semantic validity theory for linked highlighting in bar charts grounded in algebraic category theory and perceptual coherence, whilst the latter implements a highly expressive nested grammar for specifying graphical arrangements (including bar chart variants) of tractable data [16].

### 5.1 Complementary contributions

GoFish is arguably the most graphically expressive grammar to date for composing novel visualizations of tractable data. However, as discussed in Section 6.1, semantic expressivity is far more important than graphical arrangement control in statistical workflows. In this light, we might reinterpret Pollock and Satyanarayan’s [16] diagnosis of inconsistencies and expressivity cliffs in GoG-inspired systems as limitations of GoG-inspired implementations reflecting different design goals – to maximize graphical arrangement coverage rather than semantic validity. In implementations that preserve Wilkinson’s coupling of statistics and graphics, these limitations are not flaws to be solved, but design goals of the system. The constraints imposed on component-wise graphical flexibility by semantically aware statistics layers (as in *ggplot2*, rather than by the grammar itself), enable fluid navigation over semantically valid and error-free visual representations of statistical perspectives on data. To this end, formalisms and implementations of semantic constraints for valid visualizations should be viewed as distinct and complementary contributions to advancements in graphical arrangement systems [16].

Existing frameworks for conceptualizing visualization *contributions* lack support for recognizing these distinct contributions. Munzner’s [15] nested model decomposes visualization contributions into domain characterization, data/task abstraction, visual encoding and interaction idioms, and algorithm. Meyer et al. [14] extends the original model with discrete design *blocks* and *guidelines*. Although semantic validity theories loosely correspond to the data/task abstraction layer, the nested framework lacks sufficient granularity to support validity considerations. Further work is required to characterize formal validity conditions, constrained visualization design spaces and systems for composing valid graphics as distinct forms of visualization knowledge.

## 6 Discussion

### 6.1 Chart typologies, authoring and design

Graphical arrangement and semantic visualization extend the distinction between 'authoring' and 'designing' in visualization systems proposed by Satyanarayan et al. [19]. They distinguish between visualization 'authoring', where users produce a specific and known chart design, and visualization 'designing' workflows, where users engage broad exploration of visualization design spaces. Domain-specific workflows often involve both 'authoring' and 'designing' visualizations. In the statistical domain, box plots, histograms, line charts and bar charts are chart types that are commonly produced or 'authored' in the course of standard statistical analysis. This naturally leads to convenient but limited chart picker tools or helper functions that produce a fixed library of charts. However, statisticians and analysts may also engage in 'designing' charts when faced with novel statistical exploration or model diagnosis problems requiring custom or bespoke visual inspection tools, as well as when producing audience-specific and context-specific communication or presentation graphics. Both of these design workflows require more flexibility than is afforded by authoring-focused tools and chart typologies. Semantic visualization systems directly support the first design use case – i.e. via safe and error-free exploration of a semantically-constrained visualization space, but can be overly viscous in the latter design workflow. Once the semantics of the underlying data can be reduced to graphically tractable data, graphical arrangement systems, by virtue of their comparatively unconstrained design space coverage, can much more easily support the graphical adjustments and iterations needed to design effective statistical communication graphics.

### 6.2 Why grammars and compositional specifications?

Semantic visualization systems are not, by definition, restricted to grammar-based interfaces or object-oriented designs. For example, interactive scientific visualization systems often involve custom semantically valid visual encodings and fluid interactivity specific to the domain and context of the underlying data. Validity conditions are arguably quite strictly enforced – for instance by ensuring every rotational view of some 3D model accurately represents the underlying data (e.g., measurements from a medical imaging device). Similarly, graphical arrangements can be specified and rendered using a wide variety of computational interfaces, from direct-manipulation chart editors, to imperative or procedural libraries.

However, as the breadth of GoG-inspired visualization contributions show, grammar-based interfaces and modular composition are common and popular design approaches to both semantic visualization, and graphical arrangement. This shared preference likely reflects strengths of grammar-based systems that are applicable to both usage scenarios – such as flexibility relative to chart pickers, generative power and expressivity, and notational economy. Prior literature on graphical grammars suggests that they are particularly suited for the compositional flexibility required for 'designing' visualizations [17]. Grammars and compositional systems both facilitate and necessitate the decomposition of rigid chart types into graphical elements, as well as the mapping of validity conditions to grammatical components and operations. The generative power offered by grammar-based approaches applied to semantic visualization goals of fidelity, fluidity and friction naturally supports *semantic expressivity*, the more constrained notion of expressivity embodied in Bartonicek et al.'s monoidality conditions [2]. Similarly, the incremental composition of graphical grammars supports the more expansive notion of *graphical expressivity* realised in graphical arrangement grammar GoFish [16].

Semantic visualization grammar systems can also be embedded within less constrained semantic systems. For example, *ggdist* and *ggdibbler* [7, 11] implement uncertainty-aware grammar components for visualizing distributions and samples as extensions of *ggplot2*, which itself is embedded within the R programming language. However, embedding within a grammatical semantic system does not guarantee a system supports semantic visualization – for example chart type libraries can and have been built on top of *ggplot2*.

### 6.3 Evaluation of Visualization Grammars

The choice of grammar-based interface also has implications for how these systems should be evaluated. The *Cognitive Dimensions of Notations (CDN)* [3, 6] provides vocabulary for describing the "usability of notational systems" and has been applied extensively to grammar-based visualization systems. Pu and Kay [17] use CDN dimensions to characterize friction in grammar-based analyst workflows; Satyanarayan et al. [19] draw on expressivity and learnability trade-offs to assess authoring systems; Krutchen et al. [9] propose quantitative metrics for grammar coverage and notational efficiency; and McNutt [13] applies CDN systematically to compare JSON-style DSLs for visualization specification.

While the CDN dimensions provide useful shared language for comparison and analysis, they are often applied in ad-hoc manner depending on the evaluation context and goals [9]. Our distinction between system types supports the selection of appropriate CDN dimensions in a more principled and consistent manner. For graphical arrangement systems, existing CDN-grounded metrics such as arrangement coverage, specification viscosity, graphical expressivity remain appropriate, and lower viscosity is a straightforward design virtue. For semantic visualization systems, relevant metrics must instead capture: the completeness of the underlying validity theory, the degree to which invalid states are made viscous or unreachable, and the fluidity with which users can navigate the constrained but correct design space. Under these principles, Pu and Kay's [18] choice to exclude CDN dimensions which "don't distinguish PGoG from the original Grammar of Graphics" when evaluating the Probabilistic Grammar of Graphics, is also an attempt to distinguish their additional contribution over the original GoG (i.e. an embedded semantic grammar for visualizing distributions specifically).

The design goals of semantic visualization systems help to clarify trade-offs and interactions between CDN dimensions. For instance, prior work has operationalized error proneness (how easy are mistakes to miss) as a near-binary metric: PGoG uses error proneness as an evaluation criterion aimed at making errors *impossible*, and McNutt [13] similarly treats error prevention as a core benefit of DSLs. However, Error Friction relaxes this binary to allow for lower viscosity (alternative specifications are easy to reach) over valid graphics and higher viscosity over semantically invalid parts of the visualization design space, thus capturing how much notational effort a user must expend to reach semantically invalid states. This relaxation reflects a tension between semantic validity and fluid navigation, whereby making invalid states unreachable is not always necessary or desirable. Flexible enforcement may be appropriate if slight violations and specification changes don't produce perceivably different or misleading visualizations. In such cases, increasing the viscosity of invalid transitions — through warnings, soft constraints, or non-default parameter choices — achieves the error-limiting goal without adding unnecessary friction to visualization authoring processes.

### 6.4 Future Directions: Best of Both?

The distinction between semantic visualization and graphical arrangement systems is often more of a spectrum than a binary classification. Visualization grammars often enforce some validity conditions but not others, and sometimes provide extension mechanisms to circumvent validation or add new semantics. This flexibility raises the question of how to support users to move between semantic and graphical visualization authoring and design within related workflows as discussed in Section 6.1. Any transition between modes or tools likely involves a deliberate compression of semantic information into graphically tractable form. It is not clear if and when semantic reduction is safe, though it may be desirable for computational efficiency and performance. Can semantic and graphical expressivity be combined within a single system, or are users better served by clearly distinct tools? Systems that validate every operation necessarily constrain the arrangement space, and systems that maximize arrangement coverage necessarily leave validity to the user. Characterizing which of these trade-offs are principled and which are merely incidental to current implementations is an important open research problem.

Semantic validity theories may also help to overcome current limitations in AI-supported visualization. Although current vision-language models can produce visual artefacts that resemble statistical graphics in form, they cannot guarantee that the underlying statistics are valid for the chosen graphical arrangement. Semantic visualization grammars can provide the missing correctness layer by guaranteeing the statistical validity of specifications, reducing the need for post hoc verification. The low viscosity afforded by natural language interfaces could also reduce the trade-off between graphical and semantic expressivity.

## 7 Conclusion

We hope the conceptual vocabulary proposed in this paper supports more productive exchange between the statistical graphics and information visualization communities, and provides a principled basis for future work on evaluation, integration, and the design of systems that navigate the tension between graphical and semantic expressivity.